\documentclass[prl,twocolumn,superscriptaddress]{revtex4}
\usepackage{amssymb,amsmath,epsfig,bm,color,slashed,mathrsfs}
\newcommand{\be}{\begin{equation}}
\newcommand{\ee}{\end{equation}}
\newcommand{\bea}{\begin{eqnarray}}
\newcommand{\eea}{\end{eqnarray}}
\newcommand{\SD}{^3S_1$-$^3D_1}
\newcommand{\vecr}{\bm r}
\newcommand{\vecR}{\bm R}
\newcommand{\ie}{{\it i.e., }}
\newcommand{\veck}{{\bm k}}
\newcommand{\vecQ}{{\bm Q}}
\definecolor{red}{rgb}{0.8,0,0}

\begin{document}

\title{Phase diagram of dilute nuclear matter:
  Unconventional pairing and the BCS-BEC crossover}

\author{Martin Stein}
\affiliation{Institute for Theoretical Physics,
  J.~W.~Goethe-University, D-60438  Frankfurt am Main, Germany}
\author{Xu-Guang Huang}
\affiliation{Center for Exploration of Energy and Matter and
Physics Department, Indiana University, Bloomington, Indiana 47408, USA}
\affiliation{Institute for Theoretical Physics,
  J.~W.~Goethe-University, D-60438  Frankfurt am Main, Germany}
\author{ Armen Sedrakian}
\affiliation{Institute for Theoretical Physics,
  J.~W.~Goethe-University, D-60438  Frankfurt am Main, Germany}
\author{John W. Clark}
\affiliation{Department of Physics, Washington University, St.~Louis, Missouri 63130, USA}

\begin{abstract}
  We report on a comprehensive study of the phase structure of cold,
  dilute nuclear matter featuring a $\SD$ condensate at non-zero
  isospin asymmetry, within wide ranges of temperatures and densities.
  We find a rich phase diagram comprising three superfluid phases,
  namely a Larkin-Ovchinnikov-Fulde-Ferrell phase, the ordinary BCS
  phase, and a heterogeneous, phase-separated BCS phase, with
  associated crossovers from the latter two phases to a homogeneous or
  phase-separated Bose-Einstein condensate of deuterons.  The phase
  diagram contains two tricritical points (one a Lifshitz point),
  which may degenerate into a single tetracritical point for some
  degree of isospin asymmetry.
\end{abstract}
\pacs{21.65.+f, 21.30.Fe, 26.60.+c}

\maketitle

%\section{Introduction}
{\it Introduction.}
Fermionic BCS superfluids, which form
loosely bound Cooper pairs at weak coupling, undergo a transition to
the Bose-Einstein condensate (BEC) state of tightly bound bosonic
dimers, once the pairing strength
increases~\cite{1980LNP...115...13L,1985JLTP...59..195N}.  This
behavior has been confirmed in experiments on cold atomic gases, where
the interactions can be manipulated via the Feshbach
mechanism~\cite{2003Natur.426..537G,2003PhRvL..91y0401Z}.  In
isospin-symmetric nuclear matter, the transition from the BCS to the
BEC state of the $\SD$ condensate may occur upon dilution of the
system, in which case the asymptotic state is a Bose-Einstein
condensate of
deuterons~\cite{1993NuPhA.551...45A,1995PhRvC..52..975B,1995ZPhyA.351..295S,2001PhRvC..63c8201L,2006PhRvC..73c5803S,2009PhRvC..79c4304M,2010PhRvC..81c4007H,2010PhRvC..82b4911J}.
Isospin asymmetry, induced by weak interactions in stellar
environments and expected in exotic nuclei, disrupts isoscalar
neutron-proton ($np$) pairing, since the mismatch in the Fermi
surfaces of protons and neutrons suppresses the pairing
correlations~\cite{2000PhRvL..84..602S}.  The standard
Nozi\`eres-Schmitt-Rink theory~\cite{1985JLTP...59..195N} of the
BCS-BEC crossover must also be modified, such that the low-density
asymptotic state becomes a gaseous mixture of neutrons and
deuterons~\cite{2001PhRvC..64f4314L}.  The $\SD$ condensates can be
important in a number of physical settings.  (i)~Low-energy heavy-ion
collisions produce large amounts of deuterons in final states as
putative fingerprints of $SD$
condensation~\cite{1995PhRvC..52..975B}. (ii)~Large nuclei may feature
spin-aligned $np$ pairs, as evidenced by recent experimental
findings~\cite{2011Natur.469...68C} on excited states in $^{92}$Pd;
moreover, exotic nuclei with extended halos provide a locus for
$n$-$p$ Cooper pairing. (iii)~Directly relevant to the parameter
ranges covered in the present study are the observations that
supernova and hot proto-neutron-star matter at sub-saturation
densities have low temperature and low-isospin asymmetry, and that the
deuteron fluid is a substantial
constituent~\cite{2010PhRvC..81a5803T,2009PhRvC..80a5805H}.

Two relevant energy scales for the problem domain under study are
provided by the shift $\delta\mu = (\mu_n - \mu_p)/2$ in the chemical
potentials $\mu_n$ and $\mu_p$ of neurons and protons from their
common value $\mu_0$ and the pairing gap $\Delta_0$ in the $\SD$ channel
at $\delta\mu=0$.  With increasing isospin symmetry, \ie as $\delta\mu$
increases from zero to values of order $\Delta_0$, a sequence of
unconventional phases may emerge.  One of these is a neutron-proton
condensate whose Cooper pairs have non-zero center-of-mass (CM)
momentum
\cite{2001PhRvC..63b5801S,2003PhRvC..67a5802M,2009PhRvC..79c4304M};
this phase is the analog of the Larkin-Ovchinnikov-Fulde-Ferrell
(LOFF) phase in electronic
superconductors~\cite{LO,1964PhRv..135..550F}. Another possibility is
phase separation into superconducting and normal components, proposed
in the context of cold atomic gases~\cite{2003PhRvL..91x7002B}.  At
large isospin asymmetry, where $\SD$ pairing is strongly suppressed, a
BCS-BEC crossover may also occur in the isotriplet $^1S_0$ pairing
channel, notably in neutron-rich systems and halo
nuclei~\cite{PhysRevC.73.044309,2007PhRvC..76f4316M,2008PhRvC..78a4306I,PhysRevC.79.054305,2009PhRvC..79e4002A,2010PhLB..683..134S,2011PhRvC..84f7301S,2012PhRvC..86a4305S}.

Our main objective is to combine the ideas of unconventional $\SD$
pairing and the BCS-BEC crossover in a model of isospin-asymmetric
nuclear matter and construct a phase diagram for superfluid nuclear
matter over wide ranges of density, temperature, and isospin
asymmetry, while also including non-BCS {pairings}.  By doing so, we
advance the computational treatment of dilute hadronic matter along
several lines.  (i) The BCS-BEC crossover in isospin-asymmetric
systems, studied previously in Ref.~\cite{2001PhRvC..64f4314L}, is
extended to include a phase with broken spatial
symmetry and a spatially symmetric but heterogeneous phase. (ii) We
extend the earlier studies
\cite{2001PhRvC..63b5801S,2003PhRvC..67a5802M,2009PhRvC..79c4304M} of
the nuclear LOFF phase to the low density regime and show that this
phase is succeeded by a less dense heterogeneous phase before a
transition to the BEC regime occurs. (iii) We provide a
treatment of a heterogeneous (phase-separated) neutron-proton
condensate in the context of $\SD$-paired nuclear matter.
 Finally, we observe that the model explored here belongs
  to the class of imbalanced fermionic systems that has received
  wide attention in the contexts of imbalanced ultracold
  fermionic gases and color superconductivity in dense, cold
  QCD~\cite{2006pfsb.book.....S}.

%\section{Theory}
{\it Theory.}
The Green's function of the superfluid, written in the Nambu-Gorkov
basis, is given by
\bea \label{props}
i\mathscr{G}_{12} =
i\left(\begin{array}{cc} G_{12}^{+} & F_{12}^{-}\\
    F_{12}^+ &  G_{12}^{-}\end{array}\right)
= \left(\begin{array}{cc} \langle\psi_1\psi_2^+\rangle
& \langle\psi_1\psi_2\rangle \\
   \langle\psi_1^+\psi_2^+\rangle  &
   \langle\psi_1^+\psi_2\rangle
\end{array}\right),
\eea where $G_{12}^{+}\equiv G^{+}_{\alpha\beta}(x_1,x_2)$, etc., $x =
(t,\vecr) $ denotes the continuous temporal-spatial variables, and
Greek indices label discrete spin and isospin variables.  Each
operator in Eq.~(\ref{props}) can be viewed as a bi-spinor, \ie
$\psi_{\alpha}=
(\psi_{n\uparrow},\psi_{n\downarrow},\psi_{p\uparrow},\psi_{p\downarrow})^T,$
where the internal variables $\uparrow, \downarrow$ label a particle's
spin, and $n,p$ its isospin.  The matrix {propagator} obeys the
familiar Dyson equation, which has the formal solution
\be \label{Dyson} \left(\mathscr{G}_{0,13}^{-1}-\Xi_{13} \right)
\mathscr{G}_{32} = \delta_{12}, \ee in terms of the matrix
{self-energy $\Xi$}, where summation and integration over repeated indices
is implicit.  Equation~(\ref{Dyson}) is advantageously transformed into
momentum space, where it becomes an algebraic equation.  For our
purposes, translational invariance cannot be assumed, so we proceed by
defining relative and CM coordinates $\tilde r =
(x_1-x_2)$ and $R = (x_1+x_2)/2$ and Fourier transforming with respect
to the relative four-coordinate and CM three-coordinate $\vecR$. The
associated relative momentum is denoted below by $k \equiv
(ik_{\nu},\veck)$ and the three-momentum of the CM is denoted by
$\vecQ$. The zero component of the vector $k$ {takes} discrete values
$k_{\nu} = (2\nu+1)\pi T $, where $\nu\in \mathbb{Z}$ and $T$ is the
temperature. Thus the Fourier image of Eq.~(\ref{Dyson}) is written as
\be
\label{Dyson2} 
\left[\mathscr{G}_0(k,\vecQ)^{-1}-\Xi(k,\vecQ) \right]\mathscr{G}
(k,\vecQ) = {\bf{1}}_{8\times 8}.  \ee The normal propagators for the
particles and {holes} are diagonal in both spaces, \ie
$(G^+,G^{-})\propto \delta_{\alpha\alpha'}$; hence the off-diagonal
elements of $\mathscr{G}_0^{-1}$ are zero.  Writing out the
nonvanishing components in the Nambu-Gorkov space explicitly, we
obtain $[\mathscr{G}_0(ik_{\nu},\veck,\vecQ)^{-1}]_{11} =
-[\mathscr{G}_0(-ik_{\nu},\veck,-\vecQ)^{-1}]_{22}=
G_{0}^{-1}(ik_{\nu},\veck,\vecQ) $, where \be \label{inverseG}
{G}_0(k)^{-1} ={\rm diag}( ik_{\nu} - \epsilon_{n\uparrow}^+, ik_{\nu}
- \epsilon_{n\downarrow}^+, ik_{\nu} - \epsilon_{p\uparrow}^+ ,
ik_{\nu} - \epsilon_{p\downarrow}^+ ) \ee with
$\epsilon_{n\uparrow}^{\pm} = \epsilon_{n\downarrow}^{\pm} = E_S
-\delta\mu\pm E_A$ and $\epsilon_{p\uparrow}^\pm =
\epsilon_{p\downarrow}^\pm = E_S+\delta\mu \pm E_A$. Here $E_S
=\left(Q^2/4+k^2\right)/2m^*-\bar\mu$ and $E_A = \veck\cdot \vecQ
/2m^*$, with $\bar\mu \equiv (\mu_n+\mu_p)/2$.  The effective mass
$m^*$ is defined in the usual fashion in terms of the normal
self-energy, bare mass $m$, and Fermi momentum $p_F$, \ie $m/m^* =
[1-(m/p) \partial_p \Xi_{11} \vert_{p=p_F}]$, with the small mismatch
between neutron and proton effective masses being neglected.
Keeping this mismatch implies changes $E_{S/A}\to
E_{S/A}(1\pm\delta_m)$ and $\delta\mu\to \delta\mu+\mu\delta_m$, where
$\delta_m = (m^*_n-m^*_p)/(m^*_n+m^*_p)\ll 1$. In 
{the} analysis  {below,}
{$\delta_m$}  {lies}
in the range $0\le \vert \delta_m\vert \le 0.06$, the upper bound
being attained for largest asymmetries and densities relevant to this study.

The quasiparticle spectra in Eq.~(\ref{inverseG}) are
  written in a general reference frame moving with the CM momentum
  $\vecQ$ with respect to a laboratory frame at rest.  The spectrum of
  quasiparticles is seen to be two-fold degenerate; \ie the SU(4)
  Wigner symmetry of the unpaired state is broken down to spin
  SU(2).  Note that this symmetry is always approximate, since the
  phase shifts in the isoscalar and isotriplet $S$-waves differ, such
  that isosinglet pairing is stronger than isotriplet pairing in bulk
  nuclear matter.

The nucleon-nucleon scattering data show that the dominant attractive 
interaction in low-density nuclear matter is the $\SD$-partial wave, 
which leads to isoscalar (neutron-proton) spin-triplet pairing. 
Accordingly, the anomalous propagators have the property
$(F^+_{12},F^-_{12})\propto (-i\sigma_y) \otimes
\tau_x$, where $\sigma_i$ and $\tau_i$ are Pauli matrices in
isospin and spin spaces. This implies that in the quasiparticle 
approximation, the self-energy $\Xi$ has only off-diagonal elements 
in the Nambu-Gorkov space.  Specifically,   
$\Xi_{12} = \Xi_{21}^{+} = i\Delta_{\alpha\beta}$, with
$\Delta_{14}= \Delta_{23} =-\Delta_{32} = - \Delta_{41} \equiv
i\Delta$, where $\Delta$ is the (scalar) pairing gap in the $\SD$ channel.
Substituting Eq.~(\ref{inverseG}) into Eq.~(\ref{Dyson}), we obtain a set
of algebraic equations whose solutions provide the ``normal'' and
anomalous Green's functions
\bea
G_{n/p}^{\pm} &=&
\frac{ik_{\nu}\pm\epsilon_{p/n}^{\mp}}{(ik_{\nu}-E^+_{\mp/\pm})(ik_{\nu}+E^-_{\pm/\mp})},\\
F_{np}^{\pm} &=&
\frac{-i\Delta}{(ik_{\nu}-E^+_{\pm})(ik_{\nu}+E^-_{\mp})},\\
F_{pn}^{\pm} &=&
\frac{i\Delta}{(ik_{\nu}-E^+_{\mp})(ik_{\nu}+E^-_{\pm})},
\eea
the four branches of the quasiparticle spectra being given by
\be
E_{r}^{a} = \sqrt{E_S^2+\Delta^2} + r\delta\mu +a E_A,
\ee
in which $a, r \in \{+,-\}$.  Analytic continuation of
these Green's functions via $ik_{\nu} \to k_0 +i0^+$ yields
their retarded counterparts. The densities of neutrons and protons in
any of the superfluid states are obtained through 
\bea\label{eq:densities}
\rho_{n/p} (\vecQ)&=& -2\int\!\!\frac{d^4k}{(2\pi)^4}
{\rm Im} [G_{n/p}^+(k,\vecQ)-G_{n/p}^-(k,\vecQ)]f(\omega),
\nonumber\\
\eea
where $k = (k_0,\veck)$ and $f(x)=1/[\exp{(x/T)}+1]$. In mean-field approximation, the
anomalous self-energy (pairing-gap) is determined by
\bea \label{eq:gap}
\Delta(\veck,\vecQ) &=&  2i\int\!\!\frac{d^4k'}{(2\pi)^4} V(\veck,\veck')
\nonumber\\
&\times &{\rm Im} [F_{np}^+(k,\vecQ)-F_{pn}^-(k,\vecQ)]f(\omega),
\eea
where $ V(\veck,\veck')$ is the neutron-proton interaction potential.
Performing a partial-wave expansion in Eq.~(\ref{eq:gap}) as well
as the integration over $k_0$, we find
\bea \label{eq:gap2}
\Delta(Q) &=& \frac{1}{2}\sum_{a,r} \int\!\!\frac{d^3k'}{(2\pi)^3}
V_{l,l'}(k,k') \nonumber\\
&\times &\frac{\Delta_{l'}(k',Q)}{2\sqrt{E_{S}(k')^2+\Delta_{l'}(k',Q)}}[1-2f(E^r_a)],
\eea
where $V_{l,l'}(k,k')$ is the interaction in the $\SD$ partial wave.
The magnitude $Q$ of the CM momentum in Eqs.~(\ref{eq:densities}) and
(\ref{eq:gap2}) is a parameter to be determined by minimizing
the free energy of the system.  For the homogeneous (but
possibly translationally noninvariant) cases it suffices to find
the minimum of the free energy of the superfluid (S) or unpaired (N) phase,
\be \label{eq:free}
F_S =  E_S-TS_S,\quad F_N =  E_N-TS_N,
\ee
where $E$ is the internal energy (statistical average of the system
Hamiltonian) and $S$ denotes the entropy.  Stability of the superfluid 
phase requires $F_S < F_N$.  Three possibilities exist for the homogeneous
phases: (i)~$Q = 0$, $\Delta \neq 0$ (BCS phase), (ii)~$Q \neq 0$, $\Delta
\neq 0$ (LOFF phase), and $\Delta = 0$, $Q = 0$ (unpaired phase).
The free energy of the heterogeneous phase (phase-separation case) 
is constructed as a linear combination of the superfluid and unpaired energies,

\be \label{eq:free_mixed}
\mathscr{F}(x,\alpha) = (1-x) F_S(\alpha = 0) + x F_N(\alpha \neq 0),
\ee
where $x$ is the filling fraction of the unpaired component and
$\alpha = (\rho_n-\rho_p)/(\rho_n+\rho_p)$ is the density asymmetry.
  In the superfluid phase (S) one has 
$\rho_n^{(S)}=\rho_p^{(S)} = \rho^{(S)}$, whereas in the unpaired phase 
(N) they are rescaled to new values $\rho_{n/p}^{(N)}$. Thus, the 
net densities of neutrons/protons per unit volume are given by 
$\rho_{n/p} = (1-x)\rho^{(S)} + x\rho_{n/p}^{(N)}$.

%\section{Results}

{\it Results}.  Equations~(\ref{eq:densities}) and (\ref{eq:gap2}) were
solved self-consistently for a pairing interaction given by the bare
nucleon-nucleon interaction in the $\SD$ partial wave, based on the
(phase-shift equivalent) Paris potential~\cite{PhysRevC.30.1822}.  The
assumed $\SD$ partial wave implies Cooper pairing in the $S=1$, $T=0$
spin-isospin channel; $^1S_0$ Cooper pairing in the $S=0$, $T=1$
channel may mix and eventually dominate the $\SD$ pairing at
asymptotically small temperatures ($T\le 0.5$ MeV) and large
asymmetries.  Use of the bare force in Eq.~(\ref{eq:gap2}) benchmarks
the phase diagram, \ie it is reproducible with any
phase-shift-equivalent interaction.  However, some regions of the
phase diagram may strongly be affected by polarization of the
medium. Studies of polarization in neutron matter exemplify the
complexity of this problem: while propagator-based methods predict
suppression of the gap, quantum Monte-Carlo methods predict gaps
closer to the BCS result obtained with the bare force (for a recent
assessment, see~\cite{2009PhRvC..80d5802G}).  Here the nuclear mean
field was modeled by a Skyrme density functional.  The
SkIII~\cite{1987PhRvC..35.1539S} and SLy4~\cite{1998NuPhA.635..231C}
parameterizations were tested with nearly identical results.

Our results for the BCS phase and BCS-BEC
crossover are consistent with earlier studies: we observe a smooth
crossover to an asymptotic state corresponding to a mixture of a
deuteron Bose condensate and a gas of excess neutrons. The transition
from BCS to BEC is established according the following criteria:
(i)~The average chemical potential $\bar \mu$ changes its sign from
positive to negative values, and (ii)~the coherence length of a Cooper
pair becomes comparable to the interparticle distance, \ie $\xi \sim
d\sim \rho^{-1/3}$ as conditions change from $\xi \gg d$ to $\xi \ll
d$.

\begin{figure}[tb]
\begin{center}
\includegraphics[width=\linewidth]{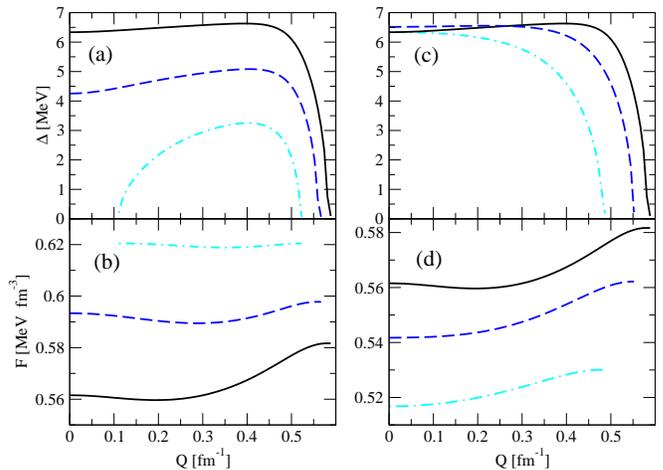}
\caption{(Color online) Properties of the nuclear LOFF phase for $\log_{10}(\rho/\rho_0)
  = -0.5$.  { Left panel:} Dependence of the pairing gap (a) and free
  energy (b) of the LOFF phase on the total momentum $\vecQ$ of a Cooper pair
  at $T=1$ MeV for asymmetries $\alpha = 0.2$ (solid, black online),
  0.3 (dashed, blue online), and 0.4 (dash-dotted, cyan online). 
  { Right panel:}  Same dependence as in the left panel, but for
  fixed asymmetry $\alpha = 0.2$ and temperatures of 1 MeV (solid, black line),
  2 MeV (dashed, blue online), and 3 MeV (dash-dotted, cyan online).  }
\label{fig:1}
\end{center}
\end{figure}
The nuclear LOFF phase arises as a result of the energetic advantage
of translational symmetry breaking by the condensate, in which pairs
acquire a nonzero CM momentum $\vecQ$.  As illustrated in
Fig.~\ref{fig:1} at $\log_{10}(\rho/\rho_0) = -0.5$, where $\rho_0=0.16$
fm$^{-3}$ is the nuclear saturation density, the gap in the LOFF phase
at nonzero asymmetries and constant temperature has its maximum at
finite $Q$, which results in a maximum of the condensation energy of
the pairs. For large asymmetries the maximum gap occurs for large
values of $Q$. At constant asymmetry, a temperature increase shifts
the gap maximum and the free-energy minimum of the LOFF phase toward
small $Q$, and at sufficiently high temperature and small asymmetry
the BCS state is favored over the LOFF phase.  This behavior is well
understood in terms of the phase-space overlap of the Fermi surfaces
of neutrons and protons, which (at finite asymmetry) increases with
temperature and the momentum $Q$ of the Cooper pairs.

Thus, as the temperature increases, we expect a restoration of the
BCS phase and of the translational symmetry in the superfluid.
Obviously, the same restoration occurs when the isospin asymmetry
is small enough.

The superfluid phase with phase separation (PS) has the symmetrical
BCS phase as one of its components. The temperature dependence of this
phase is well established within BCS theory. The second component,
which accommodates the neutron excess, is a normal Fermi liquid whose
low-temperature thermodynamics is controlled by the excitations in
the narrow strip of width $T/\epsilon_{F,n/p}$ around the Fermi surfaces
of neutrons and protons.

The transition to the BEC regime of strongly-coupled neutron-proton
pairs, which are asymptotically identical with deuterons, occurs at
low densities. As already well established, in the case of neutron-proton
pairing the criteria for the BCS-BEC transition are fulfilled, \ie
$\bar \mu$ changes sign and the mean distance between the pairs 
becomes larger than the coherence length of the superfluid.
\begin{figure}[tb]
\begin{center}
\includegraphics[width=\linewidth]{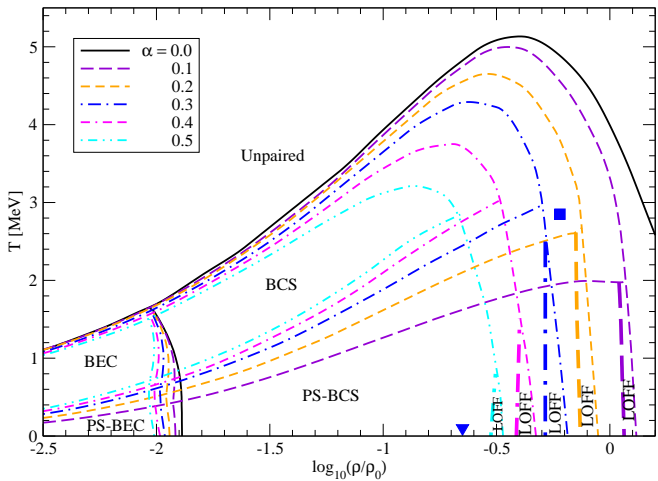}
\caption{(Color online) Phase diagram of dilute nuclear matter in the
  temperature-density plane for isospin asymmetries $\alpha = 0$
  (solid, black online), 
0.1 (dashed, violet online),
0.2 (short-dashed, orange online),
0.3 (dash-dotted, blue online),
0.4 (double-dash-dotted, magenta online), and 
0.5 (dash-double-dotted, cyan online).
  Included are four phases: unpaired, BCS (BEC) , LOFF, 
  and PS-BCS (PS-BEC). For each asymmetry there are two tri-critical 
  points, one of which is always a Lifshitz point. For special values of 
  asymmetry these two points degenerate into a single tetracritical point at
  $\log_{10}(\rho/\rho_0) = -0.22$ and $T = 2.85$ MeV (shown by a square
  dot) for $\alpha_4 = 0.255$. The LOFF phase disappears at the point
  $\log_{10}(\rho/\rho_0) = -0.65$ and $T=0$ (shown by a triangle) for
  $\alpha = 0.62.$ The density-temperature strips where the LOFF phase is
  the ground state are marked, for each asymmetry, by ``LOFF''.
}
\label{fig:2}
\end{center}
\end{figure}

We now turn to the question of how the BCS-BEC crossover is affected
by the existence of nuclear LOFF and PS phases at nonzero isospin
asymmetries, and conversely how these phases evolve in the
strongly-coupled regime if the density of the system is decreased.
The phase diagram of pair-correlated nuclear matter in the density and
temperature plane is shown in Fig.~\ref{fig:2} for several isospin
asymmetries. Four different phases of matter are present in the
diagram: (i) The unpaired phase is always the ground state of matter
at sufficiently high temperatures $T>T_{c0}$, where $T_{c0}(\rho)$ is
the critical temperature of the superfluid phase transition at $\alpha
=0$. (ii) The LOFF phase is the ground state in a narrow
temperature-density strip at low temperatures and high densities
(marked by LOFF in Fig.~\ref{fig:2}). (iii) The PS phase appears at low
temperatures and low densities. (iv)  The isospin-asymmetric BCS phase
is the ground state for all densities at intermediate temperatures. In
the extreme low-density and strong-coupling regime the BCS superfluid
phases have two counterparts. The BCS phase evolves into the BEC phase
of deuterons, whereas the PS-BCS phase evolves into the PS-BEC phase,
in which the superfluid fraction of matter is a BEC of deuterons.  The
superfluid-unpaired phase transitions and the phase transitions
between the superfluid phases are of second order (thin lines in
Fig.~\ref{fig:2}), with the exception of the PS-BCS to LOFF
transition, which is of first order (thick lines in
Fig.~\ref{fig:2}). The BCS-BEC transition and the PS-BCS to PS-BEC
transition are smooth crossovers.  At nonzero isospin asymmetry the
phase diagram features two tri-critical points where the simpler
pairwise phase coexistence terminates and three different phases
coexist.  (We do not include the points associated with crossovers
from strong to weak coupling in the class of critical points, since
these transitions involve essentially the same phase, i.e., no symmetry
is broken).

The topology of the phase diagram and the location of the tri-critical
points depends on the value of asymmetry parameter. For $\alpha
<\alpha_{4}$ the low-density critical point corresponds to coexistence
of BCS, PS, and LOFF phases, whereas the high-density critical point
corresponds to coexistence of LOFF, BCS, and unpaired phases and is
thus a Lifshitz point~\cite{Chaikin95}.  For $\alpha >\alpha_{4}$ the
topology of the phase diagram changes: The low-density tri-critical
point contains BCS, PS, and unpaired phases, whereas the high-density
tri-critical Lifshitz point contains the LOFF-PS--unpaired triple of
phases. Clearly, the point with $\log_{10}(\rho/\rho_0) = -0.22$, $T
=2.85$ MeV, and $\alpha_4 = 0.255$ is the special case of a
tetra-critical point, where all four phases; (\ie BCS, PS, LOFF, and
unpaired)  coexist.

The extreme low-density region of the phase diagram features two
crossovers. At intermediate temperatures we recover the well-known
BCS-BEC crossover, where the neutron-proton BCS condensate transforms
smoothly into a BEC gas of deuterons with some excess of neutrons.
The new ingredient of our phase diagram is the second crossover at low
temperatures, where the heterogeneous superfluid phase is replaced by
a heterogeneous mixture of a phase containing a deuteron condensate and
a phase containing neutron-rich unpaired nuclear matter. 

In closing, we note that dilute nuclear matter will definitely
  feature some clusters of higher mass number, notably $^3$He,
  $^3$H and $^4$He, coexisting in statistical equilibrium with
  the  constituents and phases revealed
  above~\cite{2010PhRvC..81a5803T,2009PhRvC..80a5805H}.  The $\alpha$
  particles ($^4$He) will form a Bose-Einstein condensate at
  sufficiently low temperatures (see Ref.~\cite{2012LNP...848..229Y}
  for a review).   These diverse aspects of superfluid,
asymmetrical nuclear matter promise significant ramifications for the
astrophysics of supernovae and (hot) compact stars and therefore
warrant examination in further detail.

This work was partially supported by the HGS-HIRe graduate program at
Frankfurt University and by the Deutsche Forschungsgemeinschaft (grant
No. SE 1836/1-2).  XGH also acknowledges support through Indiana University
grant 22-308-47 and  US Department of Energy grant No. DE-FG02-87ER40365.  AS thanks the
Kavli Institute for its Theoretical Physics China (Beijing) for hospitality. JWC
acknowledges the hospitality of the Center of Mathematical
Sciences, University of Madeira.

\end{document}